\let\includefigures=\iftrue
\let\useblackboard=\iftrue
\newfam\black
\input harvmac

\noblackbox

\includefigures
\message{If you do not have epsf.tex (to include figures),}
\message{change the option at the top of the tex file.}
\input epsf
\def\figin{\epsfcheck\figin}\def\figins{\epsfcheck\figins}
\def\epsfcheck{\ifx\epsfbox\UnDeFiNeD
\message{(NO epsf.tex, FIGURES WILL BE IGNORED)}
\gdef\figin##1{\vskip2in}\gdef\figins##1{\hskip.5in}
\else\message{(FIGURES WILL BE INCLUDED)}%
\gdef\figin##1{##1}\gdef\figins##1{##1}\fi}
\def\DefWarn#1{}
\def\figinsert{\goodbreak\midinsert}
\def\ifig#1#2#3{\DefWarn#1\xdef#1{fig.~\the\figno}
\writedef{#1\leftbracket fig.\noexpand~\the\figno}%
\figinsert\figin{\centerline{#3}}\medskip\centerline{\vbox{
\baselineskip12pt\advance\hsize by -1truein
\noindent\footnotefont{\bf Fig.~\the\figno:} #2}}
\bigskip\endinsert\global\advance\figno by1}
\else
\def\ifig#1#2#3{\xdef#1{fig.~\the\figno}
\writedef{#1\leftbracket fig.\noexpand~\the\figno}%
\global\advance\figno by1}
\fi
%

\useblackboard
\message{If you do not have msbm (blackboard bold) fonts,}
\message{change the option at the top of the tex file.}
\font\blackboard=msbm10 scaled \magstep1
\font\blackboards=msbm7
\font\blackboardss=msbm5
\textfont\black=\blackboard
\scriptfont\black=\blackboards
\scriptscriptfont\black=\blackboardss

\else

\fi
%
\def\yboxit#1#2{\vbox{\hrule height #1 \hbox{\vrule width #1
\vbox{#2}\vrule width #1 }\hrule height #1 }}
\def\fillbox#1{\hbox to #1{\vbox to #1{\vfil}\hfil}}
\def\ybox{{\lower 1.3pt \yboxit{0.4pt}{\fillbox{8pt}}\hskip-0.2pt}}

\noblackbox

\includefigures
\message{If you do not have epsf.tex (to include figures),}
\message{change the option at the top of the tex file.}
\input epsf
\def\figin{\epsfcheck\figin}\def\figins{\epsfcheck\figins}
\def\epsfcheck{\ifx\epsfbox\UnDeFiNeD
\message{(NO epsf.tex, FIGURES WILL BE IGNORED)}
\gdef\figin##1{\vskip2in}\gdef\figins##1{\hskip.5in}
\else\message{(FIGURES WILL BE INCLUDED)}%
\gdef\figin##1{##1}\gdef\figins##1{##1}\fi}
\def\DefWarn#1{}
\def\figinsert{\goodbreak\midinsert}
\def\ifig#1#2#3{\DefWarn#1\xdef#1{fig. \the\figno}
\writedef{#1\leftbracket fig.\noexpand \the\figno}%
\figinsert\figin{\centerline{#3}}\medskip\centerline{\vbox{
\baselineskip12pt\advance\hsize by -1truein
\noindent\footnotefont{\bf Fig. \the\figno:} #2}}
\bigskip\endinsert\global\advance\figno by1}
\else
\def\ifig#1#2#3{\xdef#1{fig. \the\figno}
\writedef{#1\leftbracket fig.\noexpand \the\figno}%
\global\advance\figno by1}
\fi
%

\useblackboard
\message{If you do not have msbm (blackboard bold) fonts,}
\message{change the option at the top of the tex file.}
\font\blackboard=msbm10 scaled \magstep1
\font\blackboards=msbm7
\font\blackboardss=msbm5
\textfont\black=\blackboard
\scriptfont\black=\blackboards
\scriptscriptfont\black=\blackboardss

\else

\fi
%
\def\yboxit#1#2{\vbox{\hrule height #1 \hbox{\vrule width #1
\vbox{#2}\vrule width #1 }\hrule height #1 }}
\def\fillbox#1{\hbox to #1{\vbox to #1{\vfil}\hfil}}
\def\ybox{{\lower 1.3pt \yboxit{0.4pt}{\fillbox{8pt}}\hskip-0.2pt}}

\def\hh{{1\over 2}}

\def\ll{_}
\def\uu{^}
\def\pp{\partial}
\def\L{\Lambda}
\def\exp#1{{\rm exp}\{#1\}}
\def\d{\delta}
\def\m{\mu}

\def\hsk{\hskip 1in}
\def\m{\mu}
\def\n{\nu}
\def\s{\sigma}

\def\G{\Gamma}
\def\g{\gamma}
\def\a{\alpha}

\def\e{\epsilon}
\def\O{\Omega}

\def\sqd{^2}

\def\hh{{1\over 2}}
\def\gg{\nabla}

\def\ee{\eqn\placeholder }

\def\th{\theta}

\def\ww{\wedge}
\def\mo{{-1}}
\def\b{\beta}

\def\llsk{\hskip .5in}
\def\lllsk{\hskip .15in}

\def\D{\Delta}

\def\ah{\hat{A}}
\def\lh{\hat{\lambda}}
\def\zh{\hat{Z}}
\def\yh{\hat{Y}}
\def\psih{\hat{\psi}}

\def\IN{N}
\def\hhh#1{\hat{#1}}
\def\um{^{-1}}
\def\lsq{\left [}
\def\rsq{\right ]}

\def\ZZ{Z\ll{2 \rm (dec.)}}

\def\apr{{\alpha^\prime}}

\def\l{\lambda}

\def\ww{\wedge}
\def\IZ{\relax\ifmmode\mathchoice
{\hbox{\cmss Z\kern-.4em Z}}{\hbox{\cmss Z\kern-.4em Z}}
{\lower.9pt\hbox{\cmsss Z\kern-.4em Z}} {\lower1.2pt\hbox{\cmsss
Z\kern-.4em Z}}\else{\cmss Z\kern-.4em Z}\fi}
\font\cmss=cmss10 \font\cmsss=cmss10 at 7pt
\def\inbar{\,\vrule height1.5ex width.4pt depth0pt}
\def\IC{{\relax\hbox{$\inbar\kern-.3em{\rm C}$}}}
\def\IQ{{\relax\hbox{$\inbar\kern-.3em{\rm Q}$}}}
\def\IP{\relax{\rm I\kern-.18em P}}

\lref\lennyed{
L. Susskind and E. Witten,
``The holographic bound in anti-de Sitter space,''
{\tt hep-th/9805114}.
}
\lref\dcon{
N. Arkani-Hamed, A. G. Cohen, D. B. Kaplan, A. Karch and L. Motl,
{\tt hep-th/0110146}.
}
\lref\dimdec{ N. Arkani-Hamed, A.G. Cohen and H. Georgi,
Phys.\ Rev.\ Lett.\  {\bf 86}, 4757 (2001)
{\tt hep-th/0104005}.
}
\lref\primer{
C. V. Johnson,
``D-brane primer,''
{\tt hep-th/0007170}.
}
\lref\opioneers{
I. R. Klebanov and L. Susskind,
``Continuum Strings From Discrete Field Theories,''
Nucl.\ Phys.\ B {\bf 309}, 175 (1988).
}
\lref\theymightbegiants{
J. M${^{\underline{\rm c}}}$Greevy, L. Susskind and N. Toumbas,
``Invasion of the giant gravitons from anti-de Sitter space,''
JHEP {\bf 0006}, 008 (2000)
{\tt hep-th/0003075}.
}
\lref\nebb{
G. T. Horowitz and A. Strominger,
``Black Strings And P-Branes,''
Nucl.\ Phys.\ B {\bf 360}, 197 (1991)
}
\lref\kleb{
I. R. Klebanov,
{\it  In *Shifman, M.A. (ed.): The many faces of the superworld* 307-331}.
}
\lref\ghm{
R. Gregory, J. A. Harvey and G. W. Moore,
``Unwinding strings and T-duality of Kaluza-Klein and H-monopoles,''
Adv.\ Theor.\ Math.\ Phys.\  {\bf 1}, 283 (1997)
{\tt hep-th/9708086}.
}

\lref\joeandmatt{
J. Polchinski and M. J. Strassler,
``The string dual of a confining four-dimensional gauge theory,''
{\tt hep-th/0003136}.
}

\lref\joeandmarianaone{
M. Grana and J. Polchinski,
``Supersymmetric three-form flux perturbations on AdS(5),''
Phys.\ Rev.\ D {\bf 63}, 026001 (2001)
{\tt hep-th/0009211}
}
\lref\joeandmarianatwo{
M. Grana and J. Polchinski,
``Gauge / gravity duals with holomorphic dilaton,''
Phys.\ Rev.\ D {\bf 65}, 126005 (2002)
{\tt hep-th/0106014}
}
\lref\juan{
J. M. Maldacena,
``The large $N$ limit of superconformal field theories and supergravity,''
Adv.\ Theor.\ Math.\ Phys.\  {\bf 2}, 231 (1998)
[Int.\ J.\ Theor.\ Phys.\  {\bf 38}, 1113 (1999)]
{\tt hep-th/9711200}.
}
\lref\juannc{
J. M. Maldacena and J. G. Russo,
``Large-N Limit Of Non-Commutative Gauge Theories,''
Class.\ Quant.\ Grav.\  {\bf 17}, 1189 (2000).
}
\lref\sunnyandaki{
A. Hashimoto and N. Itzhaki,
``Non-commutative Yang-Mills and the AdS/CFT correspondence,''
Phys.\ Lett.\ B {\bf 465}, 142 (1999)
{\tt hep-th/9907166}.
}
\lref\littlestring{
O. Aharony, M. Berkooz, D. Kutasov and N. Seiberg,
``Linear dilatons, NS5-branes and holography,''
JHEP {\bf 9810}, 004 (1998)
{hep-th/9808149}.
}
\lref\dm{
M. R. Douglas and G. W. Moore,
``D-branes, Quivers, and ALE Instantons,''
{\tt hep-th/9603167}.
}
\lref\dgm{
M. R. Douglas, B. R. Greene and D. R. Morrison,
``Orbifold resolution by D-branes,''
Nucl.\ Phys.\ B {\bf 506}, 84 (1997)
{\tt hep-th/9704151}.
}
\lref\kaplan{
D. B. Kaplan, E. Katz and M. Unsal,
``Supersymmetry on a spatial lattice,''
{\tt hep-lat/0206019}.
}
\lref\nima{
N. Arkani-Hamed, talk given at \it Strings 2002\rm .
}
\Title{\vbox{\baselineskip12pt\hbox{hep-th/0207226}
\hbox{SU-ITP-02/31}}}
{\vbox{
\centerline{Lattice Gauge Theories}
\bigskip
\centerline{have}
\bigskip
\centerline{Gravitational Duals}
}}
\bigskip
\bigskip
\centerline{Simeon Hellerman$^{1,2}$}
\bigskip
\centerline{$^{1}${\it Department of Physics, Stanford University,
Stanford, CA 94305}}
\smallskip
\centerline{$^{2}${\it SLAC Theory Group, MS 81, PO Box 4349,
Stanford, CA 94309}}
\bigskip
\bigskip
\noindent
In this paper we examine a certain
threebrane solution of type IIB string theory
whose long-wavelength dynamics are those of a supersymmetric
gauge theory in 2+1 continuous
and 1 discrete dimension, all of infinite extent.  Low-energy processes
in this background are described by dimensional deconstruction,
a strict limit
in which gravity decouples but the lattice spacing stays finite.
Relating this limit to the near-horizon limit of
our solution we obtain an exact, continuum
gravitational dual of a lattice
gauge theory with nonzero lattice spacing.
$H$-flux in this translationally invariant background
encodes the spatial discreteness of the gauge theory, and we
relate the cutoff on
allowed momenta to a giant graviton effect in the bulk.

\bigskip
\Date{July 24, 2002}

\newsec{Introduction}

The modern approach to string theory rests on two conceptual foundations.

The first
is that string theories in Minkowski space can be derived
as limits of conventional quantum theories as the number of degrees
of freedom per unit volume becomes large.
The second foundation is that one need not take a strict limit in order
to obtain a theory with a gravitational interpretation.  Many different
gauge theories whose large-n limits yield string or
M theory in ten or eleven flat dimensions have finite-n versions
which correspond to backgrounds with mutually distinct geometries.  The
size scales of these geometries grow as positive powers of the number
of degrees of freedom per site -- which typically means that
the size grows as a positive power of the rank $n$ of a gauge group. 

These two foundations, now established beyond dispute, lead one to wonder
whether all quantum theories with large numbers of degrees of freedom
might
have
some kind of string-theoretic interpretation and, if not, what the
criterion
could be for a nongravitational theory to have a gravitational dual.
These important questions will not be answered here.  Rather, we propose
to enlarge dramatically the class of quantum theories which admit
gravitational duals, perhaps enlarging that class so much as to plant
in the reader's mind some doubt that there may be any Hamiltonian it does
not
contain.

Most gravity/gauge theory dualities proposed so far have had two common
and related features.  On the nongravitational side we have
local quantum field theories, forumulated
in the continuum without a cutoff.\foot{Exceptions to this pattern are
proposed dualities for noncommutative quantum field theories, tensionless
string theories, and 'little
string theories' (\sunnyandaki, \juan,
\juannc, \littlestring). In this discussion we
would like to focus
on quantum theories which
one knows how to regulate and define independently of
string theory.}  On the gravitational side,
asymptotically $AdS$ geometries encode
the fact that the nongravitational description has conformal symmetry in
the
ultraviolet, which some would consider the definitive criterion
of a quantum field theory.
Asymptotically $AdS_{D+1}$ boundary conditions also express
the condition that the entropy is extensive on the $D$-dimensional
boundary
but not in the bulk, as illustrated in \lennyed, in accordance with
the holographic principle.

The second foundation of string theory expressed as above leads inevitably
to a certain question.  There is a point of view from which
gauge theories in the continuum, even at finite $n$, should themselves
be considered as limits.  Namely they like all QFT's are limits of
theories regulated with an ultraviolet cutoff at an energy scale $\L$, which is
later taken to $\infty$.  If the cutoff is a consistent, unitary quantum
regulator such as a Hamiltonian lattice theory with spacing $\L\uu{-1}$,
one is led to ask: does the regulated theory with \it finite \rm cutoff
admit its own consistent gravitational interpretation?

We answer in the  affirmative, making the
following points:

$\bullet{}$ Lattice gauge theories with
finite spacing have exact, continuum gravitational duals.

$\bullet{}$ Worldsheet instantons encode the gauge theory's
spatial discreteness, in the
form of momentum-nonconserving, or \it umklapp\rm , processes.

$\bullet{}$ The maximal size of a giant graviton in the bulk imposes an
upper bound on the
momenta of composite particles in the gauge theory.

The specific case we consider is the case of $D=4, N=4$ super-Yang-Mills
theory with one of the three spatial dimensions discretized, which breaks the
supersymmetry down by half.  We obtain the gauge theory, and the associated
supergravity solution, by starting with
a D-brane configuration in type II string theory whose long-wavelength
dynamics
are described by this discretized gauge theory, and then taking a limit
in which gravity decouples but the lattice spacing stays finite.
We note also that at infinite 't Hooft coupling, our work provides
a precise realization of the ideas of \opioneers.  At finite 't Hooft
coupling, the picture of \opioneers $ $ is corrected due to the
presence of the flux and negative spatial curvature in the
bulk.

After reviewing some background in section two,
we perform the decoupling limit in section three; in section four we identify
the stringy processes which encode momentum nonconservation; in section five
we find that the upper bound on the momentum of composit states is enforced
by a 'giant graviton' effect in the bulk; and in section six we discuss
possible applications and directions for further study.

\newsec{Dimensional deconstruction is $T$-duality}

We begin with
$n$ D2-branes near the fixed point of a $\IZ_k$ orbifold of $\IC^2$.
First we review, along the lines of \dimdec, \dcon $ $ the low-energy
dynamics of the brane sector of this theory and show that they are those
of a gauge theory on a finite periodic lattice.

We discuss scales and couplings in the theory, and two
independent limits one can
take,
the first of which corresponds to infinite volume in the lattice gauge
theory, and the second of which corresponds to the limit in which
gravity decouples from the gauge theory.  In
this section we will discuss only the first of the two.
We will also use $T$-duality to construct an
equivalent background of type IIB string theory.

\subsec{D2-branes probing orbifolds}

We now study the theory of $n$ $D2$-branes near, but not coincident with,
an $A_k$ orbifold singularity in string theory.  First we consider
branes probing the covering space.

The low-energy behavior
of a set of coincident twobranes in flat space at weak string coupling is
described by a $2+1$-dimensional gauge theory with sixteen supercharges,
which
has a unique renormalizable action.  Decomposing the matter into
multiplets under
$\IN = 4$ SUSY in 3 dimensions, we have a gauge multiplet and
a single hyperultiplet in the adjoint representation of the gauge group
$U(nk)$.  The
vector multiplet contains
a gauge field $\ah\ll\m$, three adjoint scalars $\zh\uu A$ and
four Majorana fermions $\lh\ll\a $.  The hypermultiplet contains
four scalars $\yh\uu i$ in the adjoint, and four Majorana fermions
$\psih\ll\a$, also in the adjoint.

The Lagrangian is
$$
g\ll{YM3}\sqd \CL = - {\rm tr} \hhh F\ll{\m\n} \hhh F\uu{\m\n}
- \hh (\gg\ll\m \zh\uu A) (\gg\uu\m \zh\uu A)
- \hh (\gg\ll\m \yh\uu i) (\gg\ll\m \yh\uu i)
$$
\ee
{
+ {1\over 4} {\rm tr} [ \zh\uu A , \zh\uu B ] \sqd
+ {1\over 4} {\rm tr} [\yh\uu i , \yh\uu j] \sqd
+ \hh {\rm tr} [\yh\uu i , \zh\uu A]\sqd
}
$$
+{\rm fermions}
$$
The orbifolded theory is obtained (\dm, \dgm) by truncating the fields of this
Lagrangian to the set invariant under the combined action of a global
rotation on the hypermultiplets:
\ee{
\left [ \matrix {\yh\ll 1 + i \yh\ll 2 \cr \yh\ll 3 + i \yh\ll 4 } \right
]
\to \left [ \matrix { \exp{{{2\pi i}\over k}} & 0 \cr 0 &
\exp{ -{{ 2\pi i}\over k}} }  \right ]
\left [ \matrix {\yh\ll 1 + i \yh\ll 2 \cr \yh\ll 3 + i \yh\ll 4 } \right
]
}
and a gauge transformation acting on the first tensor factor of
the gauge indices:
\ee{
\yh\ll{pp^\prime | qq^\prime}\to {\rm exp} \left\{
{{2\pi i (p - q)}\over{ k}} \right \}
\yh\ll{pp^\prime | qq^\prime} 
}
\ifig\quiver{
Quiver for branes probing a
$\IZ\ll k$ orbifold of $\IC\uu 2$ (here $k = 8$).  The circles
repreresent $U(n)$ gauge groups and the double-ended arrows
represent hypermultiplets in the $(n,\bar{n})$
of adjacent gauge groups.  There is also an eight-supercharge
vector multiplet for each gauge group, which we have not shown
explicitly.}
{\epsfxsize2.0in\epsfbox{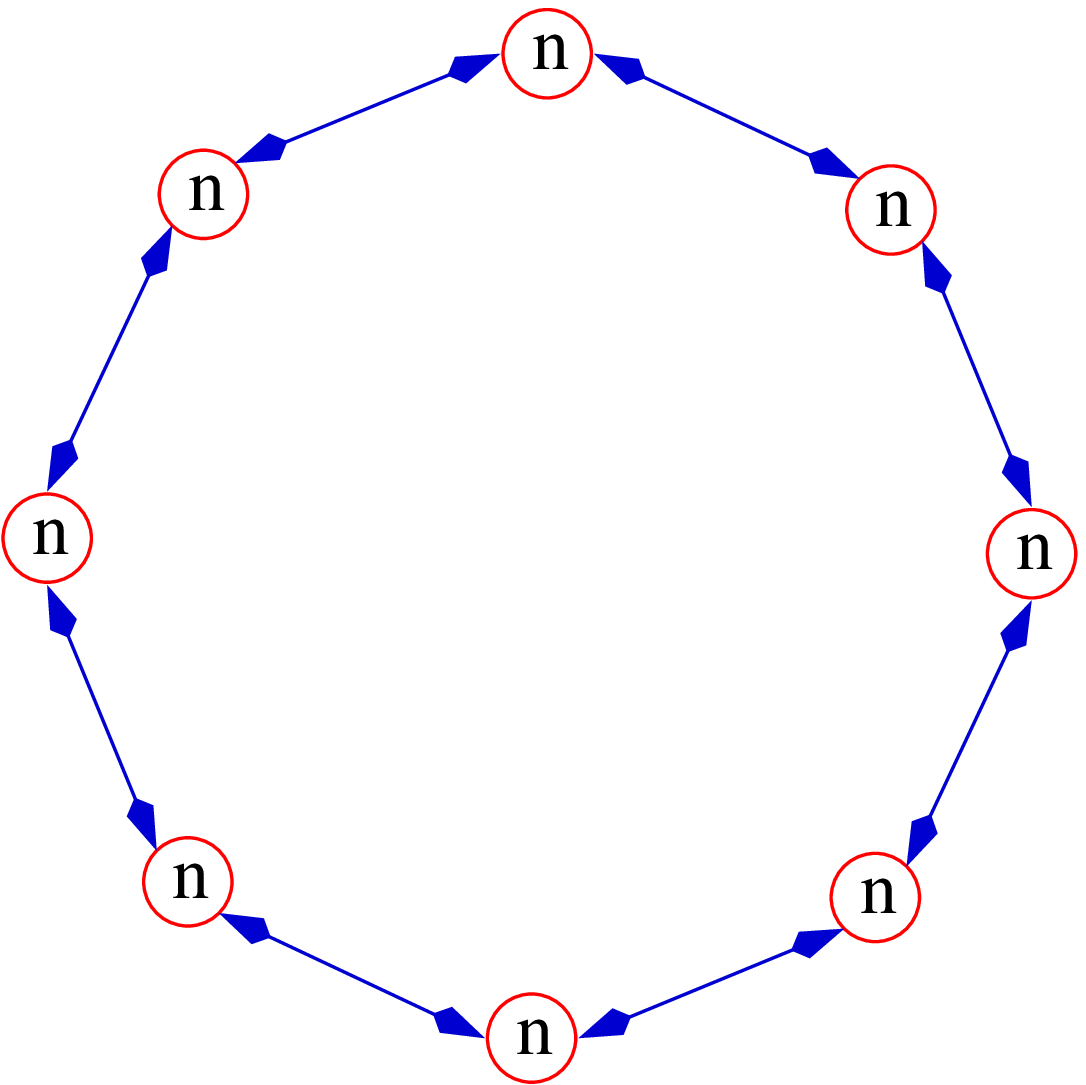}}

The components of the vector multiplets that survive the truncation
are the blocks on the diagonal:
\ee{
\zh\uu A \ll{pp^\prime | q q^\prime} = \left \{ \matrix {
0, & p\neq q \cr z\uu A \ll {p\uu\prime q\uu\prime}(p), & p=q } 
\right \}
}
$$
\ah\ll{\m,pp^\prime | q q^\prime} = \left \{ \matrix {
0, & p\neq q \cr A \ll {\m, p\uu\prime q\uu\prime}(p), & p=q } 
\right \}
$$
That is, they are adjoints of individual
$U(n)$ factors of $U(n)\uu k$.  The surviving hypers are
bifundamentals under adjacent $U(n)$'s; that is,
\ee{
\yh\uu 6 + i \yh\uu 7 \ll {pp^\prime | q q^\prime} = \left \{ \matrix {
0, & p\neq q+1 \cr (y\uu 6 + i y\uu 7) \ll
{p\uu\prime q\uu\prime}(p), & p=q+1 } 
\right \}
}
and similarly for $\yh\uu 8 - i \yh\uu 9$.  The surviving matter content
is summed up in the quiver diagram in figure 1.

If we Higgs the gauge group down to $U(n)$ by giving the
hypers a vev
\ee{
(y\uu 6 + i y\uu 7) \ll
{p\uu\prime q\uu\prime}(p) = v\ll 1 \d\ll{p\uu\prime q\uu\prime } 
}
\ee{
(y\uu 8 + i y\uu 9) \ll
{p\uu\prime q\uu\prime}(p) = v\ll 2 \d\ll{p\uu\prime q\uu\prime } ,
}
then the spectrum of massive fluctuations about the vacuum is
\ee{
E\ll j = v \sin (\pi j / k), \hsk j = 0,\cdots , k-1
}
with $v\equiv (|v\ll 1|\sqd + |v\ll 2|\sqd)\uu\hh$.
These fluctuations correspond to the lightest twisted open strings
on the D2-branes, T-dual to discrete momentum eigenmodes
on the discretized type IIB D3-brane.

The construction of the discretized threebrane from a D2-brane
probe of an orbifold proceeds in exact parallel to the discussion
of \dcon.   Our theory is just a dimensional reduction of theirs
at weak coupling, along one of the ordinary continuous dimensions.

\subsec{Scales and couplings of the probe theory}

Ultimately we will wish to take two
logically independent limits, corresponding to two
hierarchical separations of mass scales: the first, between the
string scale and the scale of the heaviest discrete momentum
modes, and the second, between the scale of the heaviest discrete
momentum modes and the scale of the lightest eigenmodes of
discrete translation, or ``Bloch waves''.

The Bloch waves on the type IIB threebrane are T-dual,
under a mod-$k$ version of the standard momentum/winding duality,
to the twisted open strings of the type IIA twobrane, and their
dispersion relation can be derived from looking at the lengths
of the type IIA twisted strings and multiplying by the string tension
$T\ll{\rm string}\equiv (2\pi\apr)\uu{-1}$.  The distance between the 
$j\uu{\underline{\rm th}}$ and $j\uu{\prime \underline{\rm th}}$ brane clump
images is just the chordal distance $2r\ll 0 |\sin \left (
{{\pi(j - j^\prime)}\over k} \right ) |$
and so the rest energy of the corresponding
twisted string is $E = {{r\ll 0}\over{\pi\apr}} |\sin \left (
{{\pi(j - j^\prime)}\over k} \right ) |$.  The dispersion relation for
Bloch waves in a free field theory with nearest neighbor kinetic terms is
\ee{
E = {\L \over \pi} \sin \left ( {{\pi |j - j^\prime|}\over{k}} \right ) 
}
for some mass scale $\L$ which we determine in terms of the lattice spacing
as follows.
The lowest nonzero mode has energy $ E\ll{\rm lowest}
\sim \L / k$.  At long distances,
the discretization is invisible, and the lowest Bloch wave looks
like a continuum fourier mode of the form $\exp{2\pi i \tilde{x}\ll 3 / V}$,
where
$V\equiv k a$ is the size of the discrete direction $\tilde{x}\ll 3$
and $a$ is
the lattice spacing.  Then $E\ll{\rm lowest} = P\ll{\rm lowest}
\sim 2\pi/ V = 2\pi / (ka)$ where
$a$ is the lattice spacing and $V \equiv ka$ is the total size of the
discrete direction.  So $\L = k /  V = 1 /  a $.  The
energy $E\ll{\rm highest}$ of the highest Bloch wave is $\L / \pi$.

Now we express these quantities in terms of string theory.
The separation between adjacent branes is approximately given by
their angular separation (on the covering space) times their
separation $r\ll 0$ from the origin.  Their angular separation is
$2\pi /k$ and so the energies of the lightest Bloch waves
go as $ E\ll{\rm lowest} = 2\pi r\ll 0 T\ll{\rm string} / k =
r\ll 0 / (k\apr) = \L / k $.  The energies
of the heaviest Bloch waves go as $ 2 r\ll 0 T\ll{\rm string}
= r\ll 0 / (\pi\apr) = \L / \pi$.  

To summarize:
$$
\L\ll{\rm string} =  {1\over{\sqrt{\apr}}}
$$
\ee{
\L \equiv \pi E\ll{\rm highest} = {r\ll 0\over{\apr}} = {{2\pi}\over a}
= {{2\pi k}\over V}
}
$$
\L\ll{\rm IR} \equiv E\ll{\rm lowest} = {r\ll 0\over{k\apr}} 
= {{2\pi}\over{ka}} = {{2\pi}\over V} = {\L\over k}
$$

The first hierarchy is therefore
\ee{
\L\ll{\rm string} / \L = \sqrt{\apr} / r\ll 0
}
and taking it to be large is the usual decoupling limit.
The second
of the two hierarchies is
\ee{
\L / \L\ll{\rm IR} = k
}
and taking this ratio to be large corresponds to the large-volume limit
of a discretized $3+1$ dimensional gauge theory in which the
dimension which is discrete is also finite finite in extent.

These two limits do not commute.
Were we to take the decoupling limit first, we would simply obtain
the sixteen supercharge gauge theory in $2+1$ dimensions whose
supergravity dual is $M$ theory on $AdS_4\times S^7$ with $n$ units
of four-form flux on the $AdS_4$.

Instead, we will first take $k$ large with the energies of the
heaviest Bloch waves held fixed.  This will yield a
$3+1$ dimensional gauge theory coupled to gravity with one
discretized dimension.  Only afterwards, once we have understood this
undecoupled lattice theory with a strictly infinite number of lattice
points will we take the decoupling limit.

\subsec{Running of the gauge coupling in the undecoupled gauge theory}

Between the string scale and the scale of the discrete momentum
modes, the nearest-neighbor kinetic terms in the discretized
direction can be ignored, and the classical scaling of the gauge
coupling is that of a $3D$ gauge theory. The effective coupling of
a $3D$ gauge theory at weak coupling runs as
$ \L \ll 1 {g\ll{YM3}\sqd}
\uu{[ \L\ll 1  ]}
=
\L\ll 2 {g\ll{YM3}\sqd}
\uu{[\L \ll 2  ]} $
plus higher perturbative corrections.
In this case we expect
 \ee{ \L {g\ll{YM3}\sqd}
\uu{[ \L ]}
=
\L\ll{\rm string} {g\ll{YM3}\sqd}
\uu{[\L \ll{\rm string} ]} .
}

The other thing to be done is to translate the 4D coupling at the lattice
scale into the 3D coupling.  The translation is simple and can be seen
by discretizing the lagrangian explicitly and restoring a canonical
normalization for the 3D fields.  This can be done in a simple
scalar model
with a $g\sqd \phi\uu 4$ coupling -- the scaling is the same.  The
relationship is simply ${g\ll 3\sqd }\uu{[\L]} = \L g\ll 4\sqd$
at the scale $\L$.  So:
\ee{
g\sqd\ll {YM4} = \L
{g\ll{YM3}\sqd}\uu{ [ \L]}
 = \L\ll{\rm string} {g\ll{YM3}\sqd}\uu{ [ \L\ll{\rm
string}]}
}
We will see later
that this matches the running of
the coupling in the string theory before
backreaction is taken into account; the spatial dependence of the
type IIB dilaton is unaffected by the backreaction of the branes.  (This
is special to the case of the discretized threebrane and does not hold
for other dimensionally deconstructed branes.)

\subsec{T-duality of the large-k orbifold}

We are first assuming we are in the 'probe' regime, in which the
backreaction of the branes on the geometry is neglected.

\ifig\orbifold{
Large$-k$ orbifold of $\IC\uu 2$.  In the diagram the branes
are represented by circles with $\times$'s in them, $\IC\uu 2$ is represented
by the plane and $k$ is represented by the number $8$.}
{\epsfxsize2.0in\epsfbox{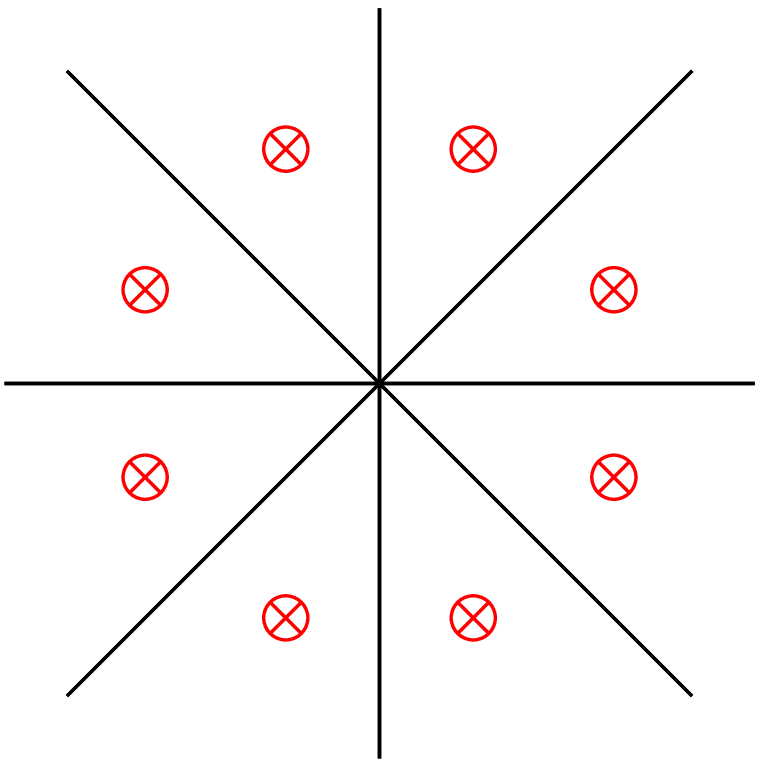}}

The metric on the orbifold is given by
$$
ds\sqd =  \eta\ll{\m\n} dx\uu\m dx\uu\n
+   dz\uu A dz\uu A
$$
\ee{
+  \left [ dr\sqd
+ {{r\sqd}\over{k\sqd}} ( d\b +
A\uu{[\b]}\ll\phi d\phi)\sqd
+ {1\over 4} r\sqd d\O\ll 2\sqd  \right ]
}

Now we $T$-dual along the $\beta$ direction $ $\primer.  If our
new, $T$-dual angular coordinate $\g$ also has periodicity
$\g\sim \g + 2\pi$, then our new metric is:
\ee{
d\tilde{s}\sqd =
\eta\ll{\m\n} dx\uu\m dx\uu\n + {{k\sqd\alpha\uu{\prime 2}}\over{r\sqd}}
d\g\sqd
+ dz\uu A dz\uu A + dr\sqd + {1\over 4}
d\O\ll 2\sqd
}
and the transformed dilaton and NS-NS two-form
are given by
\ee{
\exp{2\tilde{\Phi}}
= {{{ k\sqd \alpha^\prime} g\ll s\sqd }\over{r\sqd }} =
{{{ \alpha^\prime} \tilde{g}\ll s\sqd }\over{r\sqd }}
}
\ee{
\tilde{B}\ll{MN} dx\uu M \ww dx\uu N  =  k \apr (1 -\cos \th)
d\phi \wedge d\g
}
where $\tilde{g}\ll s \equiv k g\ll s $.

Finally we make a change of variables
\ee{
\tilde{x}\ll 3 = k \g / \L
 = \g / \L\ll{\rm IR} = k \gamma \apr / r\ll 0  
}
so
that the coordinate length $V\equiv 2\pi / \L\ll{\rm IR}$
of $\tilde{x}\ll 3$ goes to infinity with $k$.
We have chosen the normalization of $\tilde{x}\ll 3$ in such a way that
the enhanced Lorentz invariance in the infrared acts on the slice of the
geometry at the location $r = r\ll 0$ of the branes
by rotating $(x\uu 0, x\uu 1, x\uu 2, \tilde{x}\ll 3)$
as a four-vector of $SO(3,1)$ in the usual way.

Note that with this choice,
for finite $k$ the periodicity of $\tilde{x}\ll 3$ is
\ee{
\tilde{x}\ll 3 \sim \tilde{x}\ll 3 + V
}
so the newly defined coordinate $\tilde{x}\ll 3$ can indeed be
identified with the discrete compact direction in the gauge theory.

In these coordinates the type IIB string frame metric,
NS-NS $B$-field, and dilaton are given by
$$
d\tilde{s}\sqd =
\eta\ll{\m\n} dx\uu\m dx\uu\n + {r\ll 0\sqd \over{r\sqd}}
d\tilde{x}\ll 3 \sqd
+ dz\uu A dz\uu A + dr\sqd + {1\over 4}r\sqd
d\O\ll 2\sqd
$$
\ee{
\tilde{B}\ll{MN} dX\uu M \ww dX\uu N  = \apr \L (1 - \cos )\th\cdot
d\phi \wedge d\tilde{x}\ll 3
}
$$
\exp{2\tilde{\Phi}}
=
{{{ \alpha^\prime} \tilde{g}\ll s\sqd }\over{r\sqd }}
$$

In terms of tilde'd quantities, the $k\to\infty$ limit is smooth.

Let us take a look at what is happening to the gauge
coupling after we
T-dual. \foot{The author apologizes for using the word 'dual' as a
verb, but this practice has become so common that more correct
substitutes such as 'perform a duality' or even 'dualize' sound
awkward to contemporary ears. (Doubly so for that suspiciously,
subversively foreign variant: 'dualise', $c.f., e.g.$ $ $
\primer.)} The dilaton in the type IIA solution varies as
$\exp{\Phi} = g\ll s = {\rm const.}$  The action of T-duality on
the dilaton is \ee{ \exp{\tilde{\Phi}} = \exp{\Phi} \cdot
{{\alpha^{\prime \hh}}\over {R\ll\g}} = g\ll s \cdot
{{\alpha^{\prime \hh} k}\over r } = \tilde{g}\ll s \cdot
{{\alpha^{\prime \hh} }\over r } } The radial position of the
threebranes is $r = r\ll 0 = {{\alpha^\prime}\over a} = \alpha^\prime
\L$.  So the 4D gauge coupling is given by \ee{
g\sqd\ll{YM4} = \exp{\tilde{\Phi}}\ll{|\ll{r=r\ll 0}} = g\ll s  \cdot
{{\alpha^{\prime \hh} k}\over v }
 = {{k g\ll s\uu{\rm{[IIA]}} a}\over {\alpha^{\prime \hh}}} =
{{k g\ll s }\over{\alpha^{\prime\hh}\L}}
}
We now wish to take $k$ large while varying $g\ll s$ in such
a way that
$g\sqd\ll{YM4}$ stays fixed.  While this is a rather artificial
exercise from the point of view of string theory, this limit is
entirely natural from the point of view of lattice gauge theory, as
we explained earlier.
To do this we eliminate $g\ll s$ in terms of $g\sqd\ll{YM4}$:
\ee{
g\ll s  \equiv {{\alpha\uu{\prime -\hh}
v g\sqd\ll{YM4}}\over k}
= {{\alpha\uu{\prime +\hh}  g\sqd\ll{YM4}}\over {ka}}
= {{\alpha\uu{\prime +\hh}  \L g\sqd\ll{YM4}}\over k}
}
$$
\tilde{g}\ll s = \sqrt{\apr} \L g\sqd\ll{YM4} = ({r\ll 0 \over{\sqrt{\apr}}})
g\ll{YM4}\sqd
$$

\newsec{Decoupling limit of stringy dimensional deconstruction}

This section is a straightforward inclusion of the backreaction
of the branes on the geometry, and a carrying out of the usual decoupling
limit which discards the asymptotic region where the backreaction can
be neglected.

We point out that the geometry of the boundary is not asymptotically
$AdS_5$ at spatial infinity.  If we characterize points of
spatial infinity as
limits of paths with $u\sqd \equiv z\uu A z\uu A +
 r\sqd  \to \infty$
with $z\uu A / r$ and $x\uu\m, \tilde{x}\ll 3$ fixed, then
the warp factor multiplying the $(d\tilde{x}\ll 3)\sqd$
grows as a different power of $u$ at inifinity than does
the warp factor
multiplying $\eta\ll{\m\n} d x\uu\m dx\uu\n$.

In addition to the horizon and the
boundary, the solution we discuss has a
singular region at $r = z\uu A = 0$
which can be interpreted neither as a threebrane
horizon nor as a boundary. Locally this singularity represents an
infinite, continuously distributed array of NS fivebrane charge.

One interesting fact about our background is that even though the
fivebranes in the supergravity solution are smeared into a continuous
distribution along the $\tilde{x}\ll 3$ direction, the breaking of
translational invariance to a discrete subgroup -- a breaking of symmetry
of which the fivebranes are ultimately the source -- is still visible
in the supergravity approximation.  The basic physics of this has been
discussed in \ghm.  The $H$-flux transmits the breaking of
translational invariance over long distances; however fields
which transmit the breaking of the discrete translational invariance
preserved by the lattice are short-ranged and
these effects cannot be seen in the smeared supergravity solution.

To say this leaves open the question of whether
the information lost by the smeared solution is necessary
for computing amplitudes in this background.  It is certainly
true that in the undecoupled solution, giving a vev to modes
living on the NS fivebranes (T-dual to the twisted sectors in the
$\IZ\ll k$ orbifold) alters the dynamics of the threebrane probe.
It seems plausible that the interaction of
the threebranes with some subset of these modes survives
the decoupling limit.  In the holographic correspondence,
these modes would then be to frozen, non-normalizable
modes corresponding to perturbations of the gauge theory Hamiltonian
which break the discrete translational invariance of
the lattice.  The novelty would be that these modes would be
strongly supported near $r = z\uu A = 0$, rather than (or as well
as) at $u\to\infty$.

\subsec{Backreaction of D2-branes at orbifolds}

First we construct the metric produced by $nk$ $D3$-branes
distributed in a symmetric configuration on the covering space.
Using the conventions of \primer, we have

$$
ds\sqd = Z\ll p \uu{-\hh} \eta\ll{\m\n}dx\uu\m dx\uu\n + Z\ll p\uu\hh
dy\uu i dy\uu i
$$
\ee{
\exp{2\Phi} = g\ll s\sqd Z\ll p\uu{{{(3-p)}\over 2}}
}
$$
C\ll{(p+1)} = (Z\ll p\uu{-1} - 1 ) g\ll s\uu{-1} dx\uu 0 \ww \cdots
\ww dx\uu p
$$
where $\m$ ranges from $0$ to $p$, and $i$ ranges from $p$ to $9$.
$Z\ll p$ is a harmonic function of the coordinates $y\uu i$ which
we compute in the appendices.

Now, restrict to the case where $p=2$ and distribute the branes
evenly over the angular direction.  For large $k$ and fixed
$r\ll 0$, this approximation
becomes very good.  The smeared solution is
$$
ds\sqd = Z\ll 2 \uu{-\hh} \eta\ll{\m\n} dx\uu\m dx\uu\n
+ Z\ll 2 \uu{+\hh} dz\uu A dz\uu A
$$
\ee{
+ Z\ll 2 \uu{+\hh} \left [ dr\sqd
+ {1\over 4} r\sqd d\O\ll 2\sqd + {1\over {k\sqd}}r\sqd ( d\b +
 A\uu{[\b]}\ll\phi d\phi)\sqd
\right ]
}
where
\ee{ A\uu{{\rm[\b]}}\ll\phi \equiv  k (1 - 
\cos\th ) , }
$$
d\O\ll 2\sqd = d\th\sqd + \sin\sqd\th d\phi\sqd,
$$
with $Z\ll 2$ computed as a superposition of the individual harmonic
factors of the twobranes.  Our coordinate system is defined in Appendix
A and the explicit form of $Z\ll 2$
is given in Appendix B.

\subsec{T-duality and decoupling limit}

After $T$-duality our new metric is:
\ee{
d\tilde{s}\sqd = Z\ll 2 \uu{-\hh} \left [
\eta\ll{\m\n} dx\uu\m dx\uu\n + {{r\ll 0\sqd}\over{r\sqd}}
d\tilde{x}\ll 3 \sqd \right ]
+ Z\ll 2 \uu{+\hh} \left [ dz\uu A dz\uu A + dr\sqd + {{r\sqd}\over 4}
d\O\ll 2\sqd \right ]
}
and the transformed dilaton, RR potential, and NS-NS two-form
are given by
$$
\exp{2\tilde{\Phi}} = {{\alpha^\prime}\over{G\ll{\g\g}}} \exp{2\Phi}
= {{{ \alpha^\prime}k\sqd
 g\ll s\sqd Z\ll 2\uu{+\hh}}\over{r\sqd Z\ll 2\uu{+\hh}}}
= {{{ \alpha^\prime} k\sqd g\ll s\sqd }\over{r\sqd}}
= {{{ \alpha^\prime} \tilde{g}\ll s\sqd }\over{r\sqd}}
= \left ( {{r\ll 0\sqd}\over{r\sqd}}\right ) g\uu 4\ll{YM4}
$$
\ee{
\tilde{B}\ll{MN} dx\uu M \ww dx\uu N  = \apr  \L(1 - \cos\th) d\phi\ww
d\tilde{x}\ll 3 
}
$$
\tilde{C}\ll{MNST} dx\uu M \ww dx\uu N \ww dx\uu S \ww
dx\uu T = \L (Z\ll 2 \um - 1) 
dx\uu 0 \wedge dx\uu 1 \wedge dx\uu 2\wedge d\tilde{x}\ll 3 
$$
As before, the large$-k$ limit, with tilde'd quantities held
fixed, is smooth.

Taking the decoupling limit amounts to discarding the asymptotic
region.  In this limit we make the replacement
\ee{
Z\ll 2 \to \ZZ \equiv
}
\ee{
{{16 \alpha\uu{\prime 2} \tilde{g}\ll s n }\over{g\ll -\uu{3/2}
g\ll +\sqd}} \left (
 4(u\sqd + r\ll 0 \sqd)
E_2 \lsq - {{8 r\ll 0  f r }\over {g\ll -}} \rsq
-  g\ll + E_1\lsq  - {{8 r\ll 0 f r }\over {g\ll -}} \rsq
\right ).
}
$E\ll{1,2}$ and $g\ll\pm$ are defined in the appendices.

We have constructed our gauge theory to flow in the
infrared to the maximally
supersymmetric $U(n)$ gauge theory in four dimensions.
Therefore it should not surprise us that the background we have
constructed contains a horizon whose near-horizon geometry is
precisely $AdS_5\times S^5$.  To see this,
simply scale towards the point $ \phi =
z\uu A = 0, \th = \d, r =  r\ll 0$.  The backreaction factor
$Z\ll 2$ behaves in this limit as
\ee{
Z\ll 2 \propto \D\uu{-4},
}
where $\D$ is the distance to the threebranes as computed with the
type IIB metric \it without \rm the backreaction included.

\newsec{Bulk signatures of the boundary's spatial discreteness}

\subsec{Worldsheet instantons as \rm umklapp \it effects}
The most noticeable feature of lattice gauge theory is, of course,
its spatial discreteness.  There is no obvious sign of this discreteness
in the gravitational background we have constructed.
The reader may object that this is because we have simply
used the wrong metric to describe our space: after all, the geometry of
the D3-NS5 system is spatially inhomogeneous in the $\tilde{x}_3$
direction,
something which the 'smeared' solution we wrote down does not accurately
reflect.  \ifig\stringmomentumloss{
A worldsheet instanton process in which a closed string
wraps a
two-sphere and loses an amount $\Delta\tilde{p}\ll 3 = \L$
of momentum in the $\tilde{x}\ll 3$ direction.  This process
gives rise to gauge theory amplitudes which violate momentum
conservation.}
{\epsfxsize3.0in\epsfbox{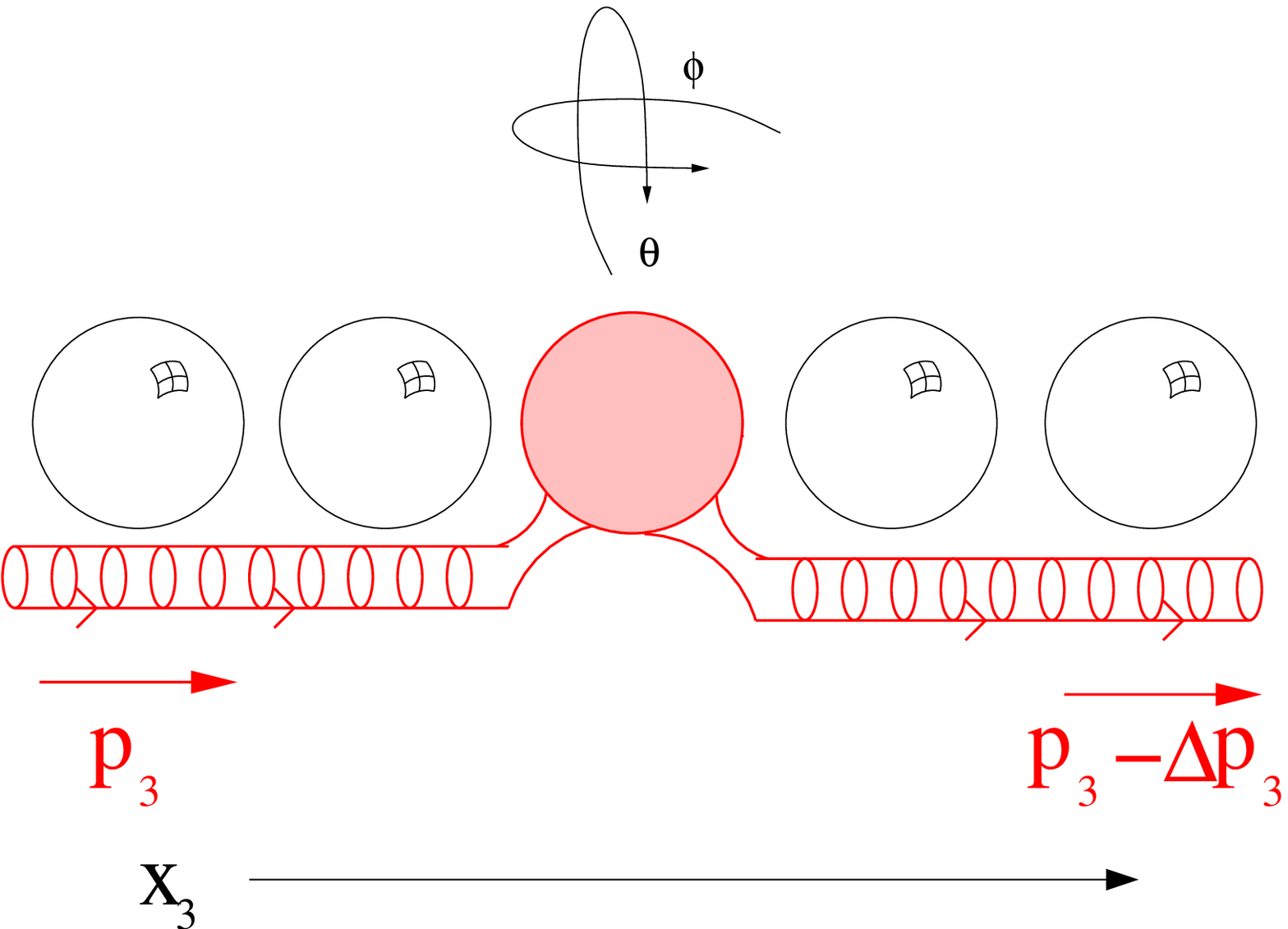}}
However
as we shall see we do \it not \rm need a solution with localized
NS fivebranes to find momentum-nonconserving processes in the
bulk theory.  Worldsheet instanton processes encode
the violation of momentum conservation in precisely the
units we expect, despite
the fact that the solution is translationally invariant.

Note that the conserved momentum of a string is not
simply its mechanical momentum
\ee{
P\ll 3  \neq P\ll{3 \rm (mech)} = {1\over{2\pi \apr}}
\oint d\s\uu 1  G\ll{\tilde{x}\ll 3 M} \dot{X}\uu M
}
The string is coupled minimally to the
NS-NS $B-$field, and the conserved quantity is the generalized momentum,
which has a term:
\ee{
 P\ll 3  = P\ll{3 \rm (mech)} +  {1\over{2\pi\apr}}
\oint d\s\uu 1 B\ll{\tilde{x}\ll 3 M}\pp\ll 1 X\uu M 
}

There is only a sensible conserved $\tilde{x}\ll 3$-momentum if the
$B$-field
is single-valued on the $S\uu 2$
and independent of $\tilde{x}\ll 3$.  For the background
we consider there is no gauge in which $B\ll{MN}$ is both.  We choose a
gauge in which $B\ll{MN}$ is independent of $\tilde{x}\ll 3$ but
not single valued.  As a result the momentum in the $\tilde{x}\ll 3$
direction will be conserved except when there are worldsheet process
which sense the non-single-valuedness of the gauge field -- that is,
momentum conservation will be violated by string instanton processes.

We now
compute the momentum loss from a worldsheet instanton in the type IIB string
background we have constructed.
The coupling of the $B$-field to the worldsheet is of the form
\ee{ {1\over{4\pi\apr}}  \int d\sqd \s \e\uu{ab} \pp\ll a X\uu\m \pp\ll
b X\uu\n }

If the $B$ field is given by $B\ll{\tilde{x}\ll 3 \phi} = -
B\ll{\phi\tilde{x}\ll 3} = \apr\L(1- \cos\th)$ then the $B$-field
contribution to the momentum $P\ll 3$ of the string is
 \ee{
P\ll 3 = P\ll {3 \rm (mech) } + {{1 \over{4\pi\apr}  }
\oint d\s\uu 1   B\ll{\tilde{x}\ll 3 \phi} } \phi^\prime  }

Fortunately the backraction factor $Z\ll 2$ does not enter into
$B\ll{MN}$.  The momentum loss is given by the integral
$$
\Delta P\ll{\tilde{x}\ll 3}
= {1\over{4\pi\apr}}  \D \oint d\s\uu 1 B\ll{\tilde{x}\ll 3 \phi}\phi
\uu\prime  
$$
$$
= {1\over{4\pi\apr}}
\int d\s\uu 0 d\s\uu 1 B\ll{\tilde{x}\ll 3 \phi,M} \phi\uu{\prime}
\dot{X}\uu M = {1\over{4\pi\apr}}
\int d\s\uu 0 d\s\uu 1 B\ll{\tilde{x}\ll 3 \phi,M} 
(\phi\uu{\prime} \dot{X}\uu M -  \dot{\phi} {X\uu M}\uu \prime)
$$
\ee{
= - {1\over{4\pi\apr}}
\int d\lsq B\ll{\tilde{x}\ll 3 M} dX\uu M \rsq
= {\L\over{4\pi}}
\int d\phi \ww d\th \sin\th = \L
}
which means the amount of $\tilde{x}\ll 3$ momentum lost by
the string is quantized in units of the inverse lattice spacing, as
one would anticipate in a theory with a lattice cutoff.

We point out that as one would expect, this process is highly suppressed
near the threebrane horizon, where instanton action diverges,
being proportional to the area of the $S\uu 2$,
which agrees with our expectation
that momentum-violating processes should be suppressed at low
energies.

\subsec{Point-particle analog}

In order better to understand this odd effect, consider the point-particle
analog in which we have an electric dipole moving on a
cylinder
in a constant magnetic field $B$.  Let the axial and angular
coordinates of the cylinder be given by $z$ and $\phi$, respectively.
We want to pick a gauge in which the gauge potential is independent
of $z$.  One such gauge is $A\ll\phi = 0$,
in which the angular momentum
of the particle has its na\"ive definition.
Then the gauge potential is $A\ll z = B \phi$.
In this gauge the momentum in the $z$ direction is not single
valued; it changes by $2\pi g B$ as the particle makes one circuit around
the $\phi$ direction.  

We can imagine a process in which the
dipole breaks apart and one of the two charged particles traverses the
circle, then binds to its partner again.  In such a process, the
momentum of the system changes by precisely $\Delta p\ll z =
\pm 2\pi g B$.  To an
observer unable to resolve the internal structure of the dipole, such
an effect would be indistinguishable from the effect of a periodic spatial
inhomogeneity of periodicity $\Delta z = {1\over{gB}}$.

How can this be? What about N\"oether's theorem?  N\"oether's theorem,
which normally 
guarantees momentum conservation in a translationally invariant system,
just never applies to the system we consider:
in a gauge in which there is a Lagrangian,
the Lagrangian depends on $z$, and in any gauge in which the gauge field
is $z$-independent, the Lagrangian does not exist as a single-valued
function on configuration space.

\ifig\particlemomentumloss{
An electric dipole in a magnetic field
can lose momentum even if the system is spatially homogeneous.}
{\epsfxsize3.0in\epsfbox{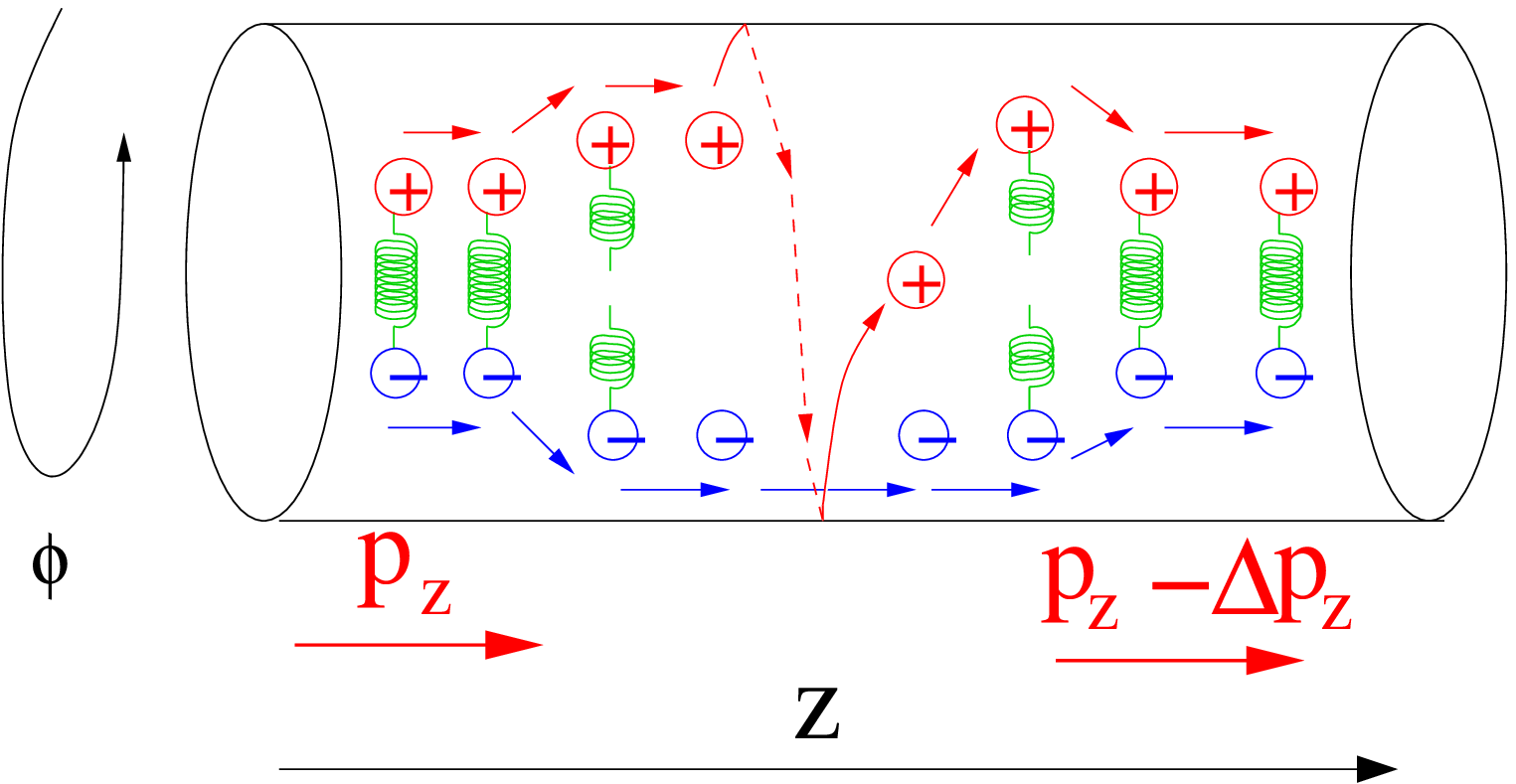}}

One can learn more by
trying to apply N\"oether's theorem to the system of dipoles coupled to
a dynamical electromagnetic field on the cylinder.  In this system, there
is no explicit breaking of translational invariance: there really is
a conserved stress tensor, the integral of
whose $T\uu 0\ll z$ component would ordinarily be a conserved momentum.
However momentum conservation is \it spontaneously \rm broken by the
presence of the background magnetic flux.  The generator of translations
contains a Poynting term $\Delta P\ll z = E\uu \th \cdot B$, and since
$B$ has a vev, the momentum has nonvanishing Poisson brackets with $A\uu\th$,
shifting the gauge field by an amount proportional to $B$.  Thus
translational invariance in this system is realized nonlinearly.
(Physically,
a charged particle which goes around the $\th$ direction leaves behind
one unit $g$ of electric flux, which contributes exactly $\mp g  B$ to
the Poynting momentum.)  The same resolution applies to our type IIB
string background.

\newsec{Giant gravitons and bounded momentum for one-particle
states}

In addition to giving rise to \it umklapp \rm processes whereby
gauge theory composites can lose momentum, the lattice also
imposes an upper bound $P\ll{\rm max}$
on composite states.  Since composites can be made up of large numbers
of partons at large 't Hooft coupling, we know that
$P\ll{\rm max}$ can in principle
be parametrically larger than $\L$. 
Is there any way to derive the bound $P\ll{\rm max}$
by considering the dual gravitational theory?

Qualitatively, the situation is rather close to that of \theymightbegiants.
The presence of the $H$-flux makes the closed fundamental string want
to blow up and wrap an $S\uu 1$ of the $S\uu 2$ parametrized by $\phi$
and $\theta$.  

In the spirit of \theymightbegiants $ $ we now estimate the
maximum momentum of
a closed fundamental string wrapping a circle of the $S\uu 2$.  The reduced
symmetry of this background makes the problem harder than the analogous
problem for giant gravitons with large angular momentum in
$AdS_p \times S^q$.  Nonetheless we find that the same basic physical
process is at work in this system as in those of \theymightbegiants.
We will find that the maximal momentum of a single particle state
is proportional to $\L$ with a coefficient of order unity, with no
powers of $n$ or the 't Hooft coupling $\l\equiv (g\sqd\ll{YM4} n)\uu\hh$;
that is, the perturbative bound for single free quanta will also apply,
up to a non-large numerical factor, to multi-parton composites in the strongly
interacting theory.

\subsec{Equations of motion}

The lagrangian on the closed string worldsheet is \ee{ S = -
{1\over{4\pi\apr}}\int d\sqd\s \left [ \sqrt{- g} g\uu{ab}
G\ll{MN}(X) +  \e\uu{ab} B\ll{MN}(X) \right ] \pp\ll a X\uu M \pp\ll b
X\uu N+ \apr\Phi(X) \cdot \sqrt{-g }R   } with the $\e$-tensor
normalized such that $\e\uu{01} = +1$.  (The action is generally
covariant because we define $d\sqd\s \equiv \hh \e\uu{ab}d\s\uu a
\wedge d\s\uu b$ so the action is invariant under coordinate
transformations which do not preserve the volume.)

The equations of motion in unit gauge $g\ll{ab} = \eta\ll{ab}$
for the embedding coordinates are then
\ee{
0 = \pp\ll a \pp\uu a X\uu S + \G\ll{MN}\uu S(\pp\ll a X\uu M)(\pp\uu a
X\uu N) - \hh {H\ll{MN}}\uu S \e\uu{ab}(\pp\ll a X\uu M)(\pp\ll b X\uu N)
}
where
\ee{
H\ll{MNS}\equiv B\ll{MN,S} + {\rm cyclic}
}

The constraint is
\ee{
0 = G\ll{MN} \pp\ll a X\uu M \pp\uu a X\uu N - \apr 
\pp\ll a \pp\ll b \Phi
}

We will now study an approximation to the actual giant
graviton problem in which the metric on the sphere is round and
its size is constant; this will capture most of the concepts
involved.

\subsec{Giant gravitons on $R\uu{1,1} \times S\uu 2$}

We examine a toy model of our problem
in which the size of the sphere is fixed,
the $H$-flux and dilaton are
constant, and the only spacetime coordinates are
$X\uu 0, X\uu 3, \phi,$ and $\th$.  We need not trouble ourselves
that such a background does not represent a solution to the spacetime
equations of motion; the only symptom is
the nonvanishing
worldsheet $\b$-function, and since we will be considering only classical
solutions this inconsistency will not impinge on our discussion.

We take the metric be
\ee{
ds\sqd = - (d x\uu 0)\sqd + (dx\uu 3)\sqd +
R\sqd [ d\th\sqd + \sin\sqd\th d\phi\sqd ]
}
and so the maximum 
where $R$ is a fixed radius that does not depend on any of the other
coordinates.  We also assume that only the azimuthal angle
$\phi$ depends on the worldsheet spatial coordinate $\s\uu 1$ and
the rest depend only on worldsheet time $\s\uu 0$.

Again taking unit gauge we find
$$
S = {1\over{4\pi\apr}} \int d\s\uu 0 d\s\uu 1 \lsq - (\dot{X}\uu 0)\sqd 
 + (\dot{X}\uu 3)\sqd + R\sqd \dot\th \sqd - R\sqd \sin\sqd \th {\phi^\prime}
\sqd + H  (1 - \cos\th) \phi^\prime \dot{X}\uu 3 \rsq
$$
\ee{
+ \hbox{terms which don't affect the solutions we'll be examining}
}

The equations of motion are:
\ee{
\dot{X} = {\rm const.} \hsk {{\pp}\over{\pp\s\uu 0}} 
(\dot{X}\uu 3 + H (1 - \cos\th) \phi\uu\prime ) = 0
}
$$
\phi\uu{\prime\prime} = 0 \hsk 
{{\pp\sqd \th }\over{(\pp\s\uu 0)\sqd}}  = \sin\th \left ( \hh H \dot{X}\uu 3
\phi^\prime - R\sqd \cos\th\phi\uu{\prime 2} \right )
$$

If we further assume that $\th$ is time-independent, then we find
\ee{
X\uu 0 =  \apr
 E \s\uu 0 \hsk X\uu 3 =  \apr E v \s\uu 0 \hsk E v = \apr P\ll 3 -
w H (1 - \cos\th)  
}
$$
\phi = w \s\uu 1 \hsk
\cos\th = {{\apr H P\ll 3 - w H\sqd }\over {2R\sqd w -  w H\sqd }}
$$
where we have allowed for an arbitrary nonzero number $w$ of windings
around the $\phi$ direction.

Since $\cos\th$ always lies between $-1$ and $+1$, this immediately
tells us that the momentum of a giant graviton must satisfy
\ee{
|P\ll 3| \leq {{2R\sqd |w|}\over{\apr |H|}}
}

So the number of momentum units per string is limited by the quantity
${{R\sqd}\over H}$.

Now we will assume that there is only a single spacetime
scale in the classical worldsheet problem.  This will
actually be true of our solution, as we shall see in the next
section.  (The string scale only controls the size of quantum corrections,
not the classical behavior of the string.)  So in our toy model we will
imagine that $H$ is not too different in size from $R$.  (In particular,
in our model $H = \apr \L = r\ll 0$, and we expect the dynamics of
the giant graviton to
prefer a point where the radius of the $S\uu 2$ is also $R \sim r\ll 0$,
since there is no other scale for it to choose.)

So we have
\ee{
P\ll 3 - \L = c \L \cdot \cos\th, 
}
where $c$ is some constant of order unity.  Depending on the actual value
of $c$, the blown-up string state may or may not exist for all possible 
values of $P\ll 3$.  But no matter what $c$ may be, the giant
solution will always exist for values of $P\ll 3$ sufficiently close
to $\L$!  So despite our lack of a solution to
the full problem the one thing we know is that near the cutoff $\L$, the
fast-moving hadron \it always \rm has a blown-up string solution.

The relative stability of the giant and pointlike states
with equal $P\ll 3$ depends
on the ratio $R / H$ and so again reduces to a problem of
classical dynamics, yet to be solved.

\subsec{Scalings in the actual giant graviton problem}

Though we have not been able to find the actual giant graviton solution
in the actual IIB background we are considering,
we can predict its coupling dependence, if it does indeed exist
(which we shall assume in this section).

We consider a nontrivial warp factors to the sphere, $X\uu 0$, and $X\uu 3$
terms in the metric; as in the actual model we consider, these warp factors
do not depend on $x\uu 0, x\uu 3,$ or $\phi$, but they may depend
on $\th$.  At this level of generality, $\th$ is distinguished from the
other coordinates only in that the $B\ll{MN}$ field depends only on
it and not on any other coordinate.  So we will denote all the other
coordinates, including $\th$, by $\{t\uu a\} = \{t\uu 0\equiv\th, t\uu 1,
t\uu 2,\cdots\}$.  

Let us now fix the dependence on the 't Hooft parameter $\l \equiv
(g\sqd n)\uu \hh$.  We will let the coefficient of the metric in the
$x\uu\m$ directions scale as $\l\um$, the coefficient of the metric
in the other directions scale as $\l\uu {+1}$, and the $B\ll{MN}$ term
scale as $\l\uu 0$, just as in our specific problem.
We have
\ee{
ds\sqd = - \l\um f\ll 1 (dx\uu 0)\sqd + \l\um f\ll 2 (dx\uu 3)\sqd 
+ \l\uu {+1} f\ll 3 d\phi\sqd
+ \l\uu{+1} h\ll{\a\b} dt\uu\a dt\uu\b
}

The worldsheet action is
\ee{
S = {1\over{4\pi\apr}} \int d\s\uu 0 d\s\uu 1 \lsq 
\l\um (- f\ll 1 (\dot{X}\uu 0)\sqd + f\ll 2 (\dot{X}\uu 3)\sqd 
)
+ H \dot{X}\uu 3 \phi^\prime (1 - \cos\th) 
+ \l\uu{+1} (h\ll{\a\b} \dot{t}\uu \a
\dot{t}\uu\b + f\ll 3 \phi^{\prime 2} )\rsq
}

Now perform the rescaling $X\uu{0,3} \equiv \l\uu{+1} \tilde{X}\uu{0,3}$.
Then the worldsheet action is:
\ee{
S = {\l\over{4\pi\apr}} \int d\s\uu 0 d\s\uu 1 \lsq 
\- f\ll 1 (\dot{\tilde{X}}\uu 0)\sqd + f\ll 2 (\dot{\tilde{X}}\uu 3)\sqd
+ H \dot{\tilde{X}}\uu 3 \phi^\prime (1 - \cos\th) 
+ h\ll{\a\b} \dot{t}\uu \a
\dot{t}\uu\b  + f\ll 3 \phi^{\prime 2}\rsq
}

The energy and momenta are defined by
\ee{
\tilde{E} = - {{\l f\ll 1}\over{\apr}}\dot{\tilde{X}}\uu 0 \hsk \tilde{P}
\ll 3  = {{\l f\ll 2}\over \apr} (\dot{\tilde{X}\uu 3} - H(1 - \cos\th)) 
}
$$
p\ll\a = {\l\over{\apr}} h\ll{\a\b} \dot{t}\uu\b
$$

The only place $\l$ appears in the action is
as an overall constant mutlitplying
all terms uniformly.
Since the problem of the existence of a giant graviton solution
is strictly a classical worldsheet problem, $\l$ drops out of the problem
completely.  Furthermore, the only spacetime
scale appearing in the classical action is $H = \alpha^\prime \L = r\ll 0$.  
Therefore on dimensional grounds alone, we can conclude
that the maximum value of $\dot{\tilde{X}}\uu 3  + \hh H \phi^\prime (
1 - \cos\th)$ is of order $r\uu 0$, and therefore $\tilde{P}\ll{\rm max}
\propto {{\l r\ll 0}\over{\apr}} = \l\L$, which means
\ee{
P\ll{\rm max}\propto \L,
} 
with constant of proportionality of order $1$.
(We have not included the varying dilaton in this discussion.  Its effect
is to change the energy of the giant graviton by altering the constraint
equation, and in principle this could effect the stability of the solution.
However the terms it contributes to $E\sqd$
scale with the same power law in $\l$
as do terms which are already included.)

This is somewhat unexpected.  Of course the maximal momentum of a
single-particle state in a weakly coupled theory should indeed scale
like $\L$, but it is far more surprising that a composite state in a
confining theory (or even a
'marginally confining' theory with vanishing beta function
in the infrared such as ours) should behave
this way.  Na\"ively, a composite with a large
number $b$ partons
in it (with $b$ of order $n$, say) should be able to have
momentum of order $b\L$.

But such a state could not be entirely stable.
Any state with momentum $>>\L$ can decay via $umklapp$'s.  The giant
graviton would represent an endpoint of many
of these $umklapp$'s in which a large number of soft
quanta share an amount of momentum of order $\L$, with
total energy of order $\L$.

It is possible that in our particular
system the giant graviton
states, like the Bloch wave excitations of the perturbative gauge theory,
are actually BPS-saturated and as a result the
momentum cutoff and dispersion relation are
exactly the same as for single quanta of the
fundamental fields of the system, namely
\ee{
E\ll{BPS} = {\L\over\pi} \sin {{|\pi P\ll 3|}\over{\L}}
}
If so, this is quite interesting, as
the bound is a non-additive BPS bound for strongly interacting
composite particles with momentum
along a direction with no continuous translational invariance!
(Other recent insights gained 
into lattice supersymmetry from the ideas of dimensional
deconstruction include \kaplan, \nima).

It would be interesting to find a
fill out a field-theoretic picture of a giant graviton state, perhaps
by computing how the momentum is shared among its many constituents.

\newsec{Conclusions}

We have constructed a gravitational dual of a theory with $2+1$ continuous
and one discrete dimension, all of infinite extent.  We have resolved
apparent paradoxes
that stem from an exact equivalence between a gauge theory in
a discrete spacetime and a gravitational theory on a continuous
background.  As in other manifestations of
gravity/gauge theory duality, one can use supergravity to learn about
the large-$n$ limits of gauge theory, or on the other hand
use the \it a priori \rm well-definedness of gauge theory to extend the
domain of definition of quantum gravity.  The construction of supergravity
duals for discretized field theories promises interesting progress along
both lines.

\subsec{The relevance of discretization}

Lattice gauge theory is a tool which is difficult to use in practice.
The lattice makes it possible to ask, and sometimes answer,
questions about
the behavior of gauge theories without appealing to a perturbative
expansion.  Relating the behavior of the lattice theory to that of
the continuum theory is difficult because the discretization at the
scale $\Lambda$ contributes corrections to the effecive action
of the continuum theory which
break Lorentz invariance and other symmetries which one wishes
to restore in the infrared.

The usefulness of the lattice theory for describing Yang-Mills theory
then rests on the irrelevance of symmetry-breaking operators.
It is not known how to compute the dimensions of these
operators in the infrared with conventional methods, since the gauge
theory is strongly coupled there.

The gravitational description, however, makes it possible to read off
the dimensions of these operators at large $n$
by computing the eigenvalues of the
Laplacian acting on linearized fluctuations of bulk fields about the
solution we have described.
This program could be carried out
not only for the $N=4, D=4$
theory but for many other strongly coupled
gauge theories if the supergravity duals of their
discretized versions can be found.

\subsec{Brane freeze: fixing the gauge coupling and lattice spacing}

One point to keep in mind is the relationship between the lattice theory
described in this paper and the type of system conventionally
referred to as 'lattice gauge theory'.  At first sight the two appear
quite different, even after allowing for the presence of the massless adjoint
matter and supersymmetry.  Our system has the additional oddity that
the $U(1)$ degree of freedom, which in the continuum theory is
completely decoupled, here couples to the branes through a nonrenormalizable
contribution to the kinetic term for the gauge fields:
\ee{
L\ll{YM4} = {1\over{4g\ll{YM4}\sqd}} 
\tr \lsq 
(1 - {{\hat{Y}\uu r }\over{2\L}}) (\hat{F\ll{\m\n}} \hat{F}\uu{\m\n}) 
\rsq + \cdots
}
where $\hat{Y}\uu r $ is a linear combination of the
hypermultiplet scalars in the gauge
theory which corresponds to infinitesimal inward motion in the $r$ direction.

We chose our decoupling limit in such a way that the
gauge coupling in the far infrared; at first sight, then, it appears
strange that the gauge coupling can be changed to any arbitrary value
by a shift, but it merely reflects the fact
that the dilaton runs as a function of the distance from the
origin $r=0$.

The same comments that apply to the gauge coupling also apply to
the lattice spacing.  If we shift $\hat{Y}\uu r $ we change the effective
lattice spacing as well.  This may appear somewhat paradoxical, since we
have shown in a previous section that the violation of conservation
of $\tilde{x}\ll 3$-momentum is strictly quantized in units of $\L$.
The resolution is that there is that the action $L\ll{YM4}$ contains
Lorentz-noninvariant terms such as
\ee{
\lsq \tr (1 - {{\hat{Y}\uu r }\over{2\L}})(\pp\ll {{\tilde{x}}\ll 3} 
\hat{X} \uu i)
(\pp\ll {{\tilde{x}}\ll 3}  \hat{X}\uu i) \rsq
}
which change the kinetic terms for all the fields in the $3$ direction
when we shift $\hat{Y}\uu r $.  The theory is still Lorentz-invariant in the
infrared, but with a different assignment of Lorentz transformations.
After we shift $\hat{Y}\uu r $, a rescaled $\tilde{x}\uu{[{\rm new}]}
\ll 3$ coordinate which enters into a four-vector with $x\uu{0,1,2}$.
The violation of $P\ll { \tilde{x}\uu{[{\rm new}]} \ll 3    }$
and the violation of $P\ll{\tilde{x}\ll 3}$ are quantized in different units,
which differ by the obvious rescaling.

In a conventional lattice gauge theory, we usually think of neither the
tree-level gauge coupling nor the lattice spacing changes as
we vary the vev of the matter fields.  But when we say the words 'lattice
gauge theory', we really mean any one of an infinite family of theories
which differ by nonrenormalizable couplings such as the very special ones
that provide the effects described above.  The theory we have been
discussing is indeed in the universality class of a conventional gauge
theory, but the presence of the massless scalars simply highlights the
fact that irrelevant terms can endow an effective theory with
dramatic effects.

Nonetheless, in order to come closer to conventional lattice gauge theory,
one may want to provide masses to some (or all) of the scalars of the
system, perhaps preserving $N=1$ or $N=2$ supersymmetry in the process.
One inviting possibility for future research is to perturb our background
with fluxes along the lines of \joeandmatt $ $ in a way which freezes
the branes in their equilibrium positions at $r = r\ll 0$.  
More precisely, the perturbation on the IIA side would look like
a $k\uu{\underline{\rm th}}$ order polynomial perturbation of the dilaton
and $p$-form fields on the
covering space of the $\IZ\ll k$ orbifold with zeroes
arranged in a ring around $r = 0$.  This solution may be obtainable
via $T$-duality to the type IIB backgrounds with holomorphic
axiodilaton described in $\joeandmarianaone, \joeandmarianatwo$.
The
combination of the superpotential
deformation with the power-law running of
the gauge coupling may provide interesting new phenomena
for the gravity/gauge theory duality to illuminate, as well as mimicking
'lattice gauge theory' (as conventionally imagined)
more closely than does the
theory we have
explored in the present paper.

\subsec{A general theory of holography?}

In the background we study we have taken the limit in which the number
of lattice sites goes to infinity, with the lattice spacing taken
to be finite.  We could have discretized all spatial directions and
taken the decoupling limit with the number of lattice sites held fixed,
still retaining a continuum gravitational description of the system
at large $n$.  It would be extremely interesting to study the
graviational background obtained this way, as it corresponds to a
system with a large but finite number of degrees of freedom per
unit volume.  (Actually one would
also have to go one step further and
impose some sort of cutoff on field space at the same time).
If one limited such a system
to a finite number of lattice sites,
the gravitational
dual would have a boundary with finite area and nonzero spatial
dimension.
Such boundaries would look locally like
Schwarzschild or deSitter horizons
and may be interesting for the study of black holes, and
also for the understanding of inflationary cosmology from a holographic
point of view.

\appendix{A}{Conventions about coordinates and indices}

Define $x\uu\m, \m = 0,1,2$ to be the longitudinal coordinates
along the type IIA D2-brane, $z\uu A, A = 7,8,9$ to be three of the transverse
coordinates, and $y\uu i, i = 3,4,5,6$ to be the four other transverse
coordinates, on which the orbifolding acts.  The radial coordinate
$r$ on $\IC\uu 2$ is defined by $r\sqd \equiv y\uu i y\uu i$
and we will sometimes use the coordinate $u$ to denote the radial
coordinate on $R\uu 7$, i.e. $u\sqd \equiv r\sqd + z\uu A z\uu Z$.
We will use capital $X\uu M$
to denote more general coordinate systems. 

Next we define four-dimensional polar
coordinates on the space $R\uu 4 =
\IC\uu 2$ spanned by the $y\uu i$:

$$
\eqalign{
\phi = \tan\uu{-1}   \left ( {{y\ll 3 y\ll 6 + y\ll 4 y\ll 5}\over{y\ll 3
y\ll 5 - y\ll 4 y\ll 6}}  \right ) \llsk & \llsk
y\ll 3 = r \cos ({\phi \over 2} + \b ) \cos (\th / 2 )
\cr
\th = \sin\uu{-1}  \left ( {{2\sqrt{y\ll 3\sqd + y\ll 4 \sqd}
\sqrt{y\ll 5 \sqd + y\ll 6 \sqd} }\over {y\ll 3 \sqd + y\ll 4 \sqd + y\ll
5
\sqd + y\ll 6 \sqd}}  \right ) \llsk
& \llsk y\ll 4 = r \sin ({\phi \over 2} + \b   ) \cos (\th / 2)
\cr
\b =  \hh\left [ \tan\uu\mo{(y\ll 4 / y\ll 3)}-
\tan\uu\mo{(y\ll 6 / y\ll 5)} \right ]\llsk & \llsk
y\ll 5 = r \cos ( {\phi \over 2} - \b   ) \sin (\th / 2) \cr
r = \sqrt{y\ll 3\sqd + y\ll 4 \sqd + y\ll 5 \sqd + y\ll 6 \sqd}
\llsk & \llsk y\ll 6 = r \sin ( {\phi \over 2} - \b   ) \sin(\th / 2)
}
$$

An element $\hat{M}$ of the
subgroup $SU(2)\ll +$ of the rotation group $SO(4) = SU(2)\ll + \times
SU(2)\ll -$ acts on these coordinates by
\ee{
\left [ \matrix { y\ll 3 + i y\ll 4 \cr y\ll 5 + i y \ll 6} \right ]
\to \hat{M}\cdot
  \left [ \matrix { y\ll 3 + i y\ll 4 \cr y\ll 5 + i y \ll 6} \right ]
}

In these coordinates the metric $G\ll{y\uu i y\uu j} = \d\uu{ij}$
on $R\uu 4 = \IC\uu 2$ is given by
\ee{
ds\sqd =  \eta\ll{\m\n} dx\uu\m dx\uu\n
+   dz\uu A dz\uu A
+  dr\sqd
+ r\sqd ( d\b + A\uu{[\b]}\ll\phi d\phi)\sqd
  + {1\over 4} r\sqd d\O\ll 2 \sqd 
}
where
\ee{
A\uu{[\b]} \ll\phi \equiv 1 - \cos\th ,
}
\ee{
d\O\ll 2\sqd = d\th\sqd + \sin\sqd\th d\phi\sqd .
}

Now, let us instead consider a metric corresponding to an a $\IZ\ll k$
orbifold, taking the same coordinate system.
The effect of the orbifolding is to reduce the proper
length of the $\b$ direction
by a factor of $k$, leaving all other metric components unchanged.
The metric on the orbifold is
\ee{
ds\sqd =  \eta\ll{\m\n} dx\uu\m dx\uu\n
+   dz\uu A dz\uu A
+  dr\sqd
+ {{r\sqd}\over{k\sqd}} ( d\b +
A\uu{[\b]}\ll\phi d\phi)\sqd
+ {1\over 4} r\sqd d\O\ll 2\sqd ,
}
where now
\ee{
A\uu{[\b]} \ll\phi \equiv k (1 -  \cos\th) .
}

\appendix{B}{Metric of the twobrane solution and its $T$-dual}

We obtain the metric
for twobranes probing the orbifold
by first taking the metric of $nk$ twobranes on the covering space,
and then taking the quotient by reducing the proper size of the
angular direction $\b$ by a factor of $k$.

To obtain the metric on the covering space, we
insert the backreaction factors $Z\ll 2\uu{-\hh}$ and
$Z\ll 2\uu{+\hh}$ into the transverse and longitudinal parts
of the metric \nebb, \primer:

\ee{
ds\sqd = Z\ll 2 \uu{-\hh} \eta\ll{\m\n} dx\uu\m dx\uu\n
+ Z\ll 2 \uu{+\hh} dz\uu A dz\uu A
}
$$
+ Z\ll 2 \uu{+\hh} \left [ dr\sqd
+ r\sqd ( d\b +
A\uu{[\b]}\ll\phi d\phi)\sqd
+ {{r\sqd\over 4}}d\O\ll 2 \sqd \right ]
$$
where
\ee{
A\uu{[\b]} \ll\phi \equiv 1 -  \cos\th
}
\ee{
Z\ll 2 \equiv 1 + \sum\ll {j= 0}\uu{k-1}
{{K\ll 2 }\over{[z\uu A z\uu A + (y\uu i -
Y\uu i\ll{(j)})\sqd ]\uu {5/2}}},
}
and where we have defined
 \ee{ K\ll 2
\equiv 6\pi\sqd g\ll s \alpha\uu{\prime (5/2)} n } and
\ee{
\left [ \matrix
{
Y\ll{(j)} \uu 3 + i Y\ll{(j)} \uu 4 
\cr
Y\ll{(j)} \uu 5 + i Y\ll{(j)} \uu 6 
}
\right ]
\equiv
\left [ \matrix { \exp { {{2\pi i j }\over k}} & 0 
\cr
0 & \exp { {{- 2\pi i j }\over k}} } \right ]
\left [ \matrix
{
Y\ll{(0)} \uu 3 + i Y\ll{(0)} \uu 4 
\cr
Y\ll{(0)} \uu 5 + i Y\ll{(0)} \uu 6 
}
\right ]
}
are the locations of the $k$ stacks of image branes.
The dilaton is
\ee{
\exp{2\Phi} = g\ll s\sqd Z\ll 2\uu{+\hh}
}
and the $C$-field is
\ee{
C\ll {(3)} = (Z\ll 2 \um - 1) dx\uu 0 \wedge dx\uu 1 \wedge dx\uu 2
}

We perform an $SO(2)$ rotation to set $Y\uu 6\ll{(0)}  = r\ll 0 \cos\d ,
Y\uu 8 \ll{(0)} =
r\ll 0\sin\d,  Y\uu 7\ll{(0)} = Y\uu 9 \ll{(0)} = 0$.
We have
\ee{
K\ll 2 \uu{-1} (Z\ll 2 - 1)  =
\sum\ll {j= 0} \uu{k-1} [ \lllsk z\uu A z\uu A +
r\sqd + r\ll 0\sqd
}
\ee{
- 2 r\ll 0r\left \{ \cos{\th\over 2}\cos{\d\over 2}\cos\left
({\phi\over 2} + \b - {{2\pi j}
\over k}\right )
+ \sin{\th\over 2}\sin{\d\over 2}
\cos\left ({\phi \over 2}- \b + {{2\pi j}
\over k}\right )\right \}
\lllsk ]\uu{-5/2}
}
\ee{
=
\sum\ll {j= 0} \uu{k-1} [ \lllsk z\uu A z\uu A +
r\sqd + r\ll 0\sqd
}
\ee{
- 2 r\ll 0r
\left \{\cos{\phi\over 2}
\cos\left ( {{\th - \d}\over 2} \right )  \cos\left ({{2\pi j}\over k}
- \b \right )
  +\sin{\phi\over 2}
 \cos\left ( {{\th + \d}\over 2}\right )  \sin\left (
{{2\pi j}\over k} - \b \right ) \right \}
\lllsk ]\uu{-5/2}
}

Because we will be considering the large-$k$ limit at fixed radius $r$,
the twobranes will become very closely spaced, so we smear out the
$D2$-brane
charge along the $\g$ coordinate.  So the branes are now marked by
continuous values $B_j \sim {{2\pi j}\over k}$ of the $\beta$-coordinate.

So then
\ee{K\ll 2 \uu{-1} (Z\ll 2 - 1)
=
\sum\ll {j= 0} \uu{k-1} [ \lllsk z\uu A z\uu A +
r\sqd + r\ll 0\sqd
}
\ee{
- 2 r\ll 0r
\left \{\cos\phi
\cos\left ( {{\th - \d}\over 2} \right )  \cos\left (B\ll j
- \b \right )
  +\sin\phi \cos\left ( {{\th + \d}\over 2}\right )  \sin\left (
B\ll j  - \b \right ) \right \}
\lllsk ]\uu{-5/2}
}

In the sum, $\D j = 1$ and $\D B =
{{2\pi}\over k}$ so
\ee{
\D B = {{\D B}\over {\D j}} = {{2\pi}\over k}
}
\ee{
\sum\ll j \to \int dj = \int {{k d B}\over {2\pi}}
}
Putting in limits of integration,
\ee{
\sum\ll{j=0}\uu{k-1} \to {k\over{2\pi}} \int\ll {B=0}\uu{B=2\pi} dB
}
Restoring the summand/integrand we have
\ee{
\sum\ll {j= 0} \uu{k-1} [ \lllsk z\uu A z\uu A +
r\sqd + r\ll 0\sqd
}
\ee{
- 2 r\ll 0r
\left \{\cos{\phi\over 2}
\cos\left ( {{\th - \d}\over 2} \right )  \cos\left (B\ll j
- \b \right )
  +\sin{\phi\over 2} \cos\left ( {{\th + \d}\over 2}\right )  \sin\left (
B\ll j  - \b \right ) \right \}
\lllsk ]\uu{-5/2}
\to
}
\ee{
{k\over{2\pi}} \int\ll 0 \uu{2\pi}
dB
 [ \lllsk z\uu A z\uu A +
r\sqd + r\ll 0\sqd
}
\ee{
- 2 r\ll 0r
\left \{\cos{\phi\over 2}
\cos\left ( {{\th - \d}\over 2} \right )  \cos (B
- \b  )
  +\sin{\phi\over 2} \cos\left ( {{\th + \d}\over 2}\right )  \sin(
B  - \b ) \right \}
\lllsk ]\uu{-5/2}
}
\ee{
= {k\over{2\pi}} \int\ll 0 \uu{2\pi}
dB
[ \lllsk z\uu A z\uu A +
r\sqd + r\ll 0\sqd
}
\ee{
- 2 r\ll 0r
\left \{\cos{\phi\over 2}
\cos\left ( {{\th - \d}\over 2} \right )  \cos B
  +\sin{\phi\over 2} \cos\left ( {{\th + \d}\over 2}\right )  \sin
B  \right \}
\lllsk ]\uu{-5/2}
}
Another shift in the variable $B$
of integration, this time by
$$
B \to B - \tan\uu{-1} \left [ \tan (\phi / 2)
\left (  {{ \cos \left ( {{\th + \d}\over 2} \right )}
\over { \cos \left ( {{\th - \d}\over 2} \right ) }} \right )  \right ],
$$
turns the integral into
\ee{
 {k\over{2\pi}} \int\ll 0 \uu{2\pi}
dB
\left [  z\uu A z\uu A +
r\sqd + r\ll 0\sqd  - 2 r\ll 0r f(\phi,\th,\d) \cos B \right ]\uu{-5/2}
}
\ee{
=
 {{2k}\over{3\pi g\ll -\uu{3/2} g\ll +\sqd }} \cdot
 \left (
 4(u\sqd + r\ll 0\sqd)
E_2 \lsq - {{8 r\ll 0 f r }\over {g\ll -}} \rsq
-  g\ll + E_1\lsq  - {{8 r\ll 0 f r }\over {g\ll -}} \rsq
\right )
}
where
\ee{f(\phi,\th,\d) \equiv \sqrt{ \cos\sqd \phi \cos\sqd \left (
{{\th - \d}\over 2} \right ) + \sin\sqd\phi \cos\sqd\left ({{\th +
\d}\over
2} \right ) },
}
\ee{
g\ll\pm \equiv
r\sqd + z\uu A z\uu A + r\ll 0\sqd \pm 4 f r\ll 0 r
= u\sqd + r\ll 0\sqd \pm 4 f r\ll 0 r,
}
and
\ee{
E\ll 1 (m) \equiv
\int\ll 0\uu {\pi \over 2} {d\a\over{ \sqrt {1 - m \sin\sqd\a}}}
}
\ee{
E_2(m) \equiv
\int\ll 0\uu {\pi \over 2} d\a \sqrt {1 - m \sin\sqd\a}
}
are complete elliptic integrals of the first and second kind.

This is what the branes look like on the covering space of the orbifold.
On the orbifold itself, the proper length of the
$\beta$-circle is reduced by a factor of $k$ relative to
its length on the covering space.
That is to say, keeping fixed the coordinate
periodicity of $\b$ at $\b\sim \b + 2\pi$,
the metric is:
\ee{
ds\sqd = Z\ll 2 \uu{-\hh} \eta\ll{\m\n} dx\uu\m dx\uu\n
+ Z\ll 2 \uu{+\hh} dz\uu A dz\uu A
}
\ee{
+ Z\ll 2 \uu{+\hh} \left [ dr\sqd
+ {1\over 4} r\sqd d\O\ll 2\sqd + {1\over {k\sqd}}r\sqd ( d\g +
A\uu{[\b]}\ll\phi d\phi)\sqd
\right ]
}
where
\ee{ A\uu{{\rm[\b]}}\ll\phi \equiv k(1 - \cos\th)   , }
\ee{
d\O\ll 2\sqd = d\th\sqd + \sin\sqd\th d\phi\sqd,
}
and
\ee{
Z\ll 2   = 1 +
{{16 \alpha\uu{\prime 2} \tilde{g}\ll s n }\over{g\ll -\uu{3/2}
g\ll +\sqd}} \left (
 4(u\sqd + r\ll 0\sqd)
E_2 \lsq - {{8 r\ll 0 f r }\over {g\ll -}} \rsq
-  g\ll + E_1\lsq  - {{8 r\ll 0 f r }\over {g\ll -}} \rsq
\right )
}

The effect of the decoupling limit is to drop the constant term $1$ in
the backreaction factor, replacing $Z\ll 2$ with
\ee{
\ZZ =
 (Z\ll 2 - 1)
=
}
\ee{
{{16 \alpha\uu{\prime 2} \tilde{g}\ll s n }\over{g\ll -\uu{3/2}
g\ll +\sqd}} \left (
 4(u\sqd + r\ll 0\sqd)
E_2 \lsq - {{8 r\ll 0 f r }\over {g\ll -}} \rsq
-  g\ll + E_1\lsq  - {{8 r\ll 0 f r }\over {g\ll -}} \rsq
\right )
}

\bigskip
\centerline{\bf{Acknowledgements}}
The author would like to thank Allan Adams, Keshav Dasgupta, John
M${^{\underline{\rm c}}}$Greevy, Joseph Polchinski,
Mohammed Sheikh-Jabbari, Stephen Shenker,
Leonard Susskind, and Scott Thomas for
valuable discussions.  I am particularly grateful to Prof. Susskind
for an explanation of the behavior of strings in discrete target spaces and
for drawing my attention to existing results on the subject.
I also thank the Harvard theory
group for hospitality while this work was in progress.
This work was supported by the DOE under contract
DE-AC03-76SF00515.






%

\listrefs
\end